# Can Smaller Large Language Models Evaluate Research Quality?

Mike Thelwall

Although both Google Gemini (1.5 Flash) and ChatGPT (4o and 4o-mini) give research quality evaluation scores that correlate positively with expert scores in nearly all fields, and more strongly that citations in most, it is not known whether this is true for smaller Large Language Models (LLMs). In response, this article assesses Google's Gemma-3-27b-it, a downloadable LLM (60Gb). The results for 104,187 articles show that Gemma-3-27b-it scores correlate positively with an expert research quality score proxy for all 34 Units of Assessment (broad fields) from the UK Research Excellence Framework 2021. The Gemma-3-27b-it correlations have 83.8% of the strength of ChatGPT 4o and 94.7% of the strength of ChatGPT 4o-mini correlations. Differently from the two larger LLMs, the Gemma-3-27b-it correlations do not increase substantially when the scores are averaged across five repetitions, its scores tend to be lower, and its reports are relatively uniform in style. Overall, the results show that research quality score estimation can be conducted by offline LLMs, so this capability is not an emergent property of the largest LLMs. Moreover, score improvement through repetition is not a universal feature of LLMs. In conclusion, although the largest LLMs still have the highest research evaluation score estimation capability, smaller ones can also be used for this task, and this can be helpful for cost saving or when secure offline processing is needed.
**Keywords**: Scientometrics; Large Language Models; Gemma; Open weights LLMs

## Introduction

Research evaluation is a common and important task for academics and managers, and it is often supported by citation-based indicators (Hicks et al., 2015; Moed, 2005; Mukherjee, 2022). With the increasingly widespread use of Artificial Intelligence (AI) in research (Mohammadi et al., 2025), it is important to check whether it can save expert time through support of the research evaluation task. ChatGPT research quality score estimates for journal articles are recent alternatives to citations as quantitative indicators to support evaluations (Kousha & Thelwall, 2025). Their value lies in their positive correlation with expert judgement in all or nearly all fields, and at a slightly higher rate than for citation-based indicators (Thelwall, 2025abc). Despite some systematic biases or disparities (Thelwall & Kurt, 2025), this property means that they are helpful when expert judgement fails, such as for areas outside of the assessor's expertise, as a cross-check for bias, and for evaluations where assessment expertise is unavailable or too expensive for the value of the task (Thelwall, 2025d).

Whilst a positive correlation with expert judgement has been established for three of the largest Large Language Models (LLMs) in 2025, ChatGPT 4o, ChatGPT 4o-mini, and Google Gemini Flash 1.5 (Thelwall, 2025ac), these are all cloud-based services and may be too expensive or not private enough for some research evaluation purposes (Nowak et al., 2025). Moreover, cloud-based services can be withdrawn, updated, or made more costly, so research evaluation procedures may not be able to rely on them. Thus, there is a need to test whether any smaller "open weights" LLMs (Sowe et al., 2024) that can be downloaded and used offline have a capability to estimate research quality.

## Literature review

LLMs are a type of artificial intelligence with a specific design (architecture) and trained on an enormous quantity of text. This training gives them a capability to generate grammatically correct and usually meaningful language in response to a user prompt. With an instruction tuning additional stage, they are also able to respond with reasonable answers to a wide variety of user requests (Chung et al., 2024). Multimodal LLMs are trained on images as well as text and can respond to prompts like, "Generate an appropriate caption for this image" (Qi et al., 2023). Another relevant concept is "mixture of experts" (Jacobs et al., 1991), the idea that a single AI system can include multiple subsystems specialising in different tasks. LLMs can therefore produce substantially different results for the same input if it is routed to a different "expert" sublayer (Shazeer et al., 2017). LLMs are inherently language-agnostic but are typically trained on multiple languages and can perform services like translation as well as responding to prompts in multiple languages (Huang et al., 2024).

One unusual phenomenon is that some LLM capabilities do not exist for smaller versions but only appear when the size is large enough: these are called emergent properties (Wei et al., 2022). The opposite is a predictable, scalable property that may be weak for smaller models but gets stronger as model size increases. Because of the potential for emergent properties, large LLM capabilities cannot be guaranteed to exist in smaller LLMs without testing. Thus, it is not clear whether the research quality scoring capability of ChatGPT and Gemini also exists for smaller LLMs.

The magnitude of an LLM can be judged by its file size or the number of parameters, when saved. LLMs can be made smaller through the process of quantisation, which retains the number of parameters but saves them at a reduced level of accuracy (Xiao et al., 2023). Overall, the capability of an LLM depends on its number of parameters, overall design (architecture) and the precision with which the parameters are stored. LLMs are available for download at various sizes from a few gigabytes to hundreds of gigabytes on the huggingface.co platform.

Although most investigations of LLMs for research quality estimation have used ChatGPT or Gemini, as mentioned in the introduction, one previous study has used locally run smaller LLMs. It used different versions of Llama3 and Quent2.5, comparing them with ChatGPT 4o-mini. The task was a three-class proxy quality rating for biomedical research. Zero-shot and fewshot were tried (and fine tuning for non-generative models). Qwen 2.5-72b and ChatGPT 4o-mini performed best, but precision and recall were reported rather than correlations so it is not possible to judge whether the smaller LLMs had a capability to rank articles for research quality, Nevertheless, the highest Cohen's kappa reported was only 0.059 (for zero-shot) suggesting a very weak ability for this (Wu et al., 2025). Positive results have been obtained from smaller LLMs for the different task of pre-publication conference paper reviewing, however (Zhou et al., 2024).

## Methods

As discussed above, the capability of smaller LLMs has not been checked directly for research quality scores and has not been checked outside of biomedicine. The objective of the current study is to this gap with a science wide study directly investigating research quality scores with the following research questions. The second question, about the need for repetitions, is relevant because previous research has shown that averaging multiple independent scores for the same article from ChatGPT or Gemini gives a higher correlation with expert scores than individual scores. This needs to be checked for smaller LLMs because repetition increase the cost of the scores and so should be avoided if unnecessary.

- RQ1: Can smaller (i.e., downloadable open weights) LLMs give research quality scores that correlate positively with expert judgement in all fields of science?
- RQ2: Does averaging LLM scores across multiple repetitions increase the correlation with expert judgement?

For RQ1, the research design was to score a large dataset of journal articles from all fields of science with the smaller LLM Gemma and then correlate the scores with an indicator of the expert quality scores of those articles. For RQ2, the average correlations with individual scores were compared with the correlations with average scores across five repetitions.

### *Journal article dataset*

As used in all previous science-wide evaluations of LLMs, this article uses journal articles without short abstracts submitted to the UK's Research Excellence Framework (REF) 2021 as the dataset to analyse. This is a large set of primary research articles (excluding reviews) published between 2014 and 2020 and selected by researchers in UK higher education institutions as their best 1-5 outputs during the period. These outputs were each given an individual score of 1* (nationally relevant), 2* (internationally relevant), 3* (internationally excellent) or 4* (world leading) by two senior researchers from the relevant broad field during a year. This is the largest scale systematic research quality evaluation ever conducted and is a serious process because the results direct the UK's annual block grant for research, which totals about £14 billion over the lifetime of REF2021. Unfortunately, the scores for individual journal articles are not released but average scores for sets of articles are public. The sets are very approximately department size (the terminology used here). REF 2021 was organised into 34 broad disciplines, called Units of Assessment (UoAs) and each university could submit the works of one group of researchers (two in some circumstances) to each UoA.

The public results for the REF are the average scores for each university and UoA (or group and UoA when more than one submission from a university to a UoA). Since the average scores vary substantially between universities, these average scores are reasonable proxies for the individual article scores. Thus, the dataset analysed here is REF2021 journal articles and the associated average departmental research quality scores. For each UoA, the articles with the 10% shortest abstracts were removed. The 10% was determined heuristically to remove articles without abstract or that were short form contributions (e.g., research notes) that are not equivalent to standard journal articles. Since the same dataset has been used before, the correlations for them are directly comparable to those reported here.

For the research quality scoring, each article is represented by its title and abstract rather than its full text. This is for two reasons. First, ChatGPT's scores from titles and abstracts give higher correlations with expert judgement than its scores from full text. Second, not all of the article full texts are available without charge online.

### *Research quality scores from Gemma*

Google Gemma-3-27b-it is a multimodal, multilingual, instruction tuned open weights LLM that can be downloaded from huggingface.co (https://huggingface.co/google/gemma-3-27b-it) with a combined file size of 59.9Gb in safetensors format and 27 billion parameters. Its total size is perhaps misleading because its image processing capabilities are not used here. Gemma is described by Google as a family of "lightweight" LLMs that it shares freely with the research community. The current version, Gemma 3, has a 128k token context window, which easily allows it to process article titles and abstracts. It is available in four sizes, based on the number of parameters, with the smallest having 1 billion and the largest 27 billion (as used here). The models were released on March 2025 (https://developers.googleblog.com/en/introducing-gemma3/).

All articles in the dataset were submitted to Gemma x for evaluation with the same system instructions as previously used for ChatGPT and Gemini. These instructions (see: Thelwall, 2025b) define the research quality evaluation task. The system instructions are slight paraphrases of the instructions given to the REF2021 expert reviewers. There were four sets of expert reviewer instructions and hence there are four corresponding system instructions. These are for the groupings of UoAs into main panels A (health and life sciences, UoAs 1-6), B (engineering and physical sciences, UoAs 7-12), C (social sciences, UoAs 13-24), and D (arts and humanities, UoAs 25-34).

Each set of system instructions explains that the task is to evaluate the quality of a journal article, defining the four different star levels, discussing the factors that might be considered when assigning a score, and defining research quality as comprising rigour, originality and significance. These are the most common dimensions for research quality definitions (Langfeldt et al., 2020).

For each evaluation, Gemma was fed with the appropriate system instructions for the article (based on its UoA) then the user prompt "Score this:" followed by the article title, a new line, the word "Abstract", another newline, and the abstract on a single line. Each score request was submitted in a separate session to avoid the score from one article influencing the score for another. This was repeated five times (non-consecutively) for each article and the arithmetic mean of the five scores used as the final score. Although Gemma is an open weights LLM, this process was conducted through the free Google Gemma API (https://ai.google.dev/gemma/docs/core/gemma_on_gemini_api).

The Gemma output from the above prompts is typically a report evaluating the article. Within the report there is almost always a score on the specified scale. This was extracted by a program for the purpose (WA link). The recommended score was always extracted except that if Gemini gave a fractional score (e.g., 2.67 and then rounded it to an exact star level (3* in this case) then the unrounded number was used.

*Correlations*

The LLM scores were correlated with the departmental average research quality scores separately for each UoA because the average scores vary between UoAs and the UoA-level results may reveal field differences in LLM capability. Spearman correlations were used because the primary use of the LLM scores is to rank articles rather than to assign a given score. For example, the quantitative indicators supplied to some REF2021 evaluators were percentile ranks rather than score predictions. Bootstrapping was used to calculate the Spearman correlation confidence intervals since there is not a formula for this.

# Results

Overall, Gemma rarely used the lowest score (1*) but still tended to give lower scores than the REF experts overall (Table 1). Gemma also tended to avoid giving any score higher than 3* for social sciences, arts and humanities fields.

Table 1. Descriptive statistics for the dataset analysed and the Gemma scores. Totals and averages are from UoAs not articles (i.e., including double counting for articles in multiple UoAs).

| UoA | Articles | REF mean | Gemma minimum | Gemma maximum | Gemma mean |
|---|---|---|---|---|---|
| 1 - Clinical Medicine | 9604 | 3.28 | 2 | 4 | 2.97 |
| 2 - Public Health, Health Services and Primary Care | 3822 | 3.34 | 2 | 4 | 2.94 |
| 3 - Allied Health Professions, Dentistry, Nursing & Pharmacy | 9340 | 3.16 | 2 | 4 | 2.77 |
| 4 - Psychology, Psychiatry and Neuroscience | 7876 | 3.11 | 2 | 4 | 2.87 |
| 5 - Biological Sciences | 6124 | 3.32 | 2 | 4 | 2.99 |
| 6 - Agriculture, Food and Veterinary Sciences | 3070 | 3.12 | 2 | 4 | 2.84 |
| 7 - Earth Systems and Environmental Sciences | 3378 | 3.34 | 2 | 4 | 2.92 |
| 8 - Chemistry | 2325 | 3.40 | 2 | 4 | 2.90 |
| 9 - Physics | 3634 | 3.39 | 2 | 4 | 2.95 |
| 10 - Mathematical Sciences | 2965 | 3.39 | 2 | 4 | 2.92 |
| 11 - Computer Science and Informatics | 4122 | 3.20 | 2 | 4 | 2.56 |
| 12 - Engineering | 14358 | 3.24 | 1 | 4 | 2.59 |
| 13 - Architecture, Built Environment and Planning | 2239 | 3.14 | 2 | 3 | 2.41 |
| 14 - Geography and Environmental Studies | 2788 | 3.11 | 2 | 3.2 | 2.79 |
| 15 - Archaeology | 426 | 3.13 | 2 | 4 | 2.82 |
| 16 - Economics and Econometrics | 702 | 3.23 | 2 | 4 | 2.82 |
| 17 - Business and Management Studies | 8939 | 3.05 | 2 | 3 | 2.55 |
| 18 - Law | 1471 | 2.99 | 2 | 4 | 2.50 |
| 19 - Politics and International Studies | 2077 | 3.01 | 2 | 3 | 2.73 |
| 20 - Social Work and Social Policy | 2724 | 2.96 | 2 | 3 | 2.49 |
| 21 - Sociology | 1266 | 2.98 | 2 | 3 | 2.63 |
| 22 - Anthropology and Development Studies | 750 | 2.99 | 2 | 4 | 2.75 |
| 23 - Education | 2852 | 2.94 | 1.8 | 3 | 2.43 |
| 24 - Sport and Exercise Sciences, Leisure and Tourism | 2423 | 3.13 | 2 | 3.6 | 2.48 |
| 25 - Area Studies | 383 | 3.07 | 2 | 3 | 2.52 |
| 26 - Modern Languages and Linguistics | 719 | 3.03 | 2 | 3 | 2.50 |
| 27 - English Language and Literature | 516 | 2.92 | 2 | 3 | 2.45 |
| 28 - History | 673 | 2.84 | 2 | 3 | 2.70 |
| 29 - Classics | 54 | 3.05 | 2 | 3 | 2.54 |
| 30 - Philosophy | 419 | 3.04 | 2 | 3 | 2.71 |
| 31 - Theology and Religious Studies | 135 | 2.71 | 2 | 3 | 2.48 |
| 32 - Art and Design: History, Practice and Theory | 811 | 2.89 | 2 | 3 | 2.25 |
| 33 - Music, Drama, Dance, Performing Arts, Film & Screen Studies | 417 | 2.85 | 1 | 3 | 2.34 |
| 34 - Comm, Cultural & Media Studies, Library & Information Management | 785 | 2.98 | 2 | 3 | 2.42 |
| Total/Average | 104,187 | 3.10 | - | - | 2.66 |

The Gemma scores correlated positively with the departmental average scores in all fields and the correlations were statistically significantly different from 0 in 30 of the 34 UoAs, with the exceptions being smaller UoAs (426 or fewer articles) (Figure 1). Thus, Gemma has a universal or near universal ability to estimate research quality scores, although this ability is weak in most fields (below 0.3) or moderate.

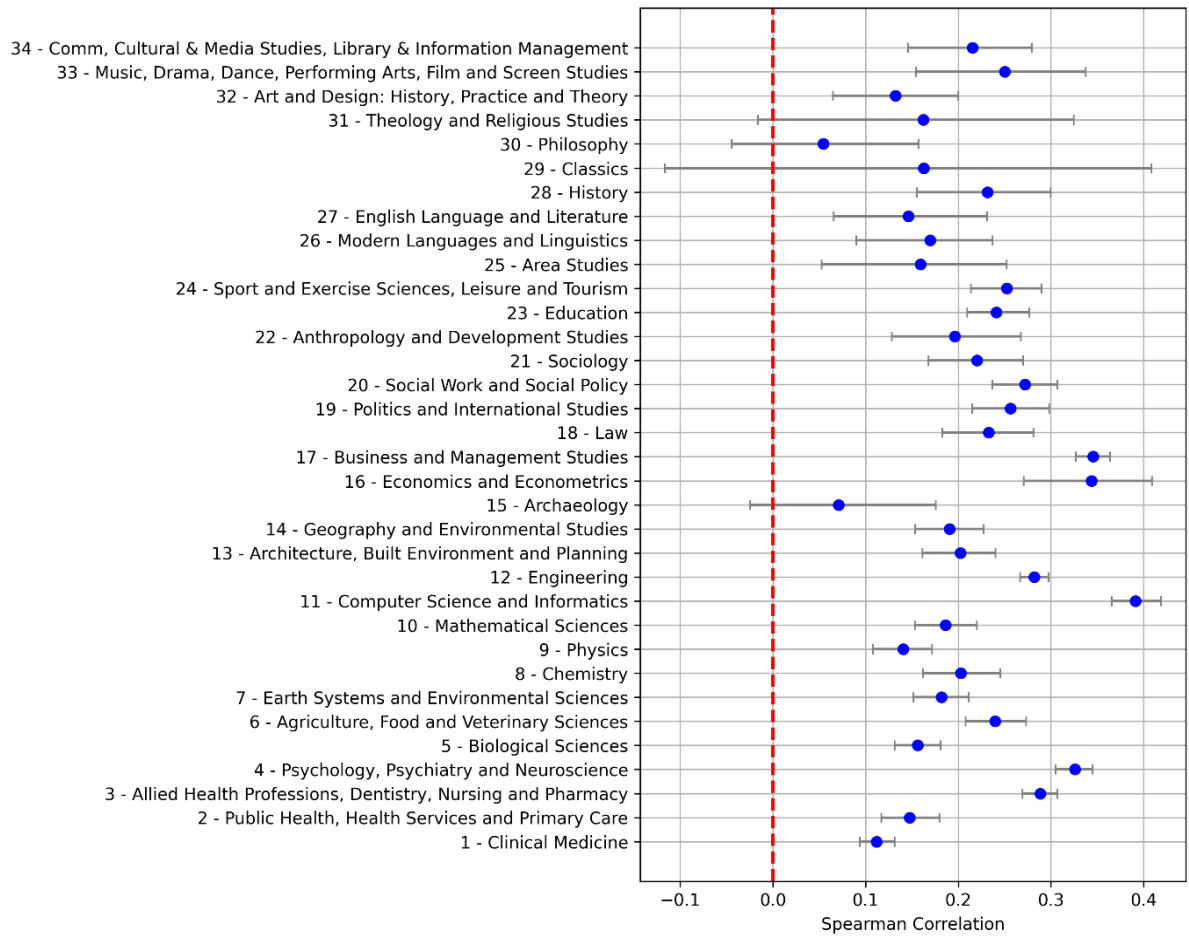

Figure 1. Spearman correlations and 95% bootstrapped confidence intervals for Gemma research quality scores against departmental average REF scores as a proxy for individual article scores, by UoA.

Averaging multiple iterations of Gemma makes little difference to the results (Figure 2). Averaging five iterations increased the correlation by 2% (Main Panels A and C), 1% (Mai Panel B) or 0.1% (Main Panel D) overall. The main reason for the lack of change was that the score rarely changed. In 95.7% of articles, all five scores were identical. In contrast, averaging 5 iterations increased the Gemini correlation by 8% overall, with only 45% of articles having identical scores all five times (author's calculation from the raw data).

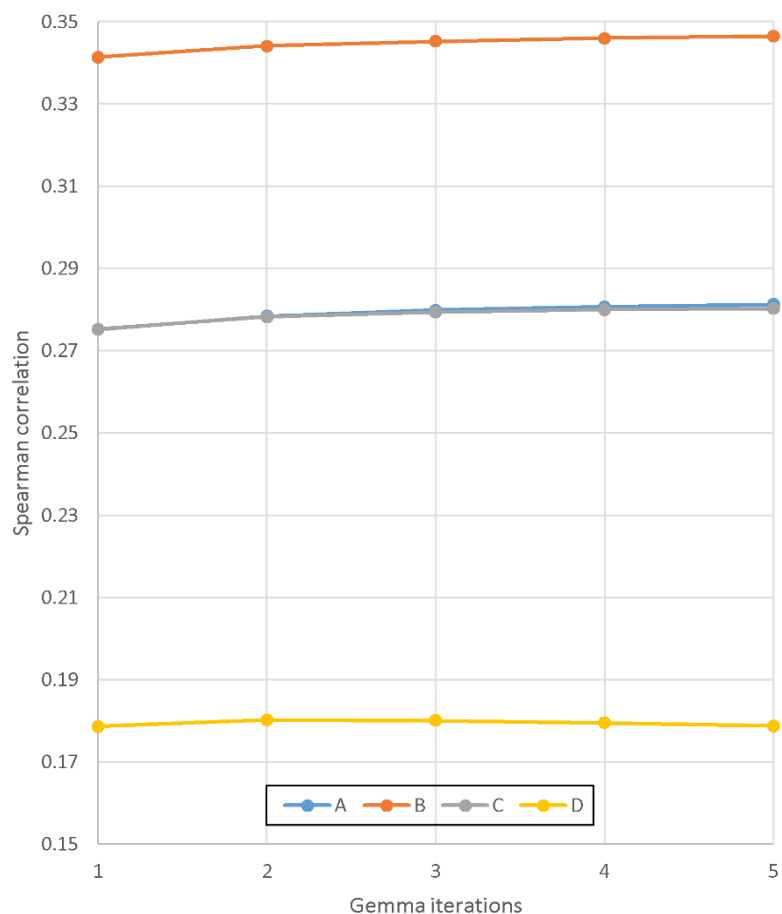

Figure 2. Spearman correlations for Gemma research quality scores against departmental average REF scores as a proxy for individual article scores for different numbers of Gemma iterations averaged, by main panel. Main panels A and C almost perfectly overlap. Note that the y axis does not start at 0.

## Discussion

This study is limited by the single UK-based dataset and the use of departmental average quality scores rather than individual article quality scores. This probably has a dampening effect on the correlations. Gemini might work less well on other countries' outputs, particularly if they are not in English. It might also have a weaker ability to estimate scores for different definitions of research quality. The results are likely to be different for other LLMs, and some may have no ability to estimate research quality scores. Finally, higher correlations may have been obtained from other similar sized LLMs, or with other strategies (e.g. fewshot, fine tuning, other system instructions).

### *Comparison with prior research*

In comparison to ChatGPT 4o and 4o-mini (as reported in: Thelwall, 2025b), the correlations with expert REF scores (using the departmental average proxy again) tend to be lower, although the difference is not large (Figure 3). In a few cases the Gemma correlations are higher, however (e.g., UoA 17). Using a sample size weighed mean, the average Spearman correlations across all means are ChatGPT 4o: 0.285, ChatGPT 4o -mini: 0.252, and Gemma-3-27b-it: 0.239. Thus, overall, the Gemma-3-27b-it correlations have 83.8% of the strength of ChatGPT 4o and 94.7% of the strength of ChatGPT 4o-mini. Surprisingly, given the broadly similar correlations, the average score estimates tended to be substantially higher for ChatGPT 4o (3.22) and ChatGPT 4o-mini (3.15) than for Gemma-3-27b-it (2.66). Unfortunately, the correlations are not directly comparable to those published before for Gemini Flash 1.5 (Thelwall, 2025a) because these were based on smaller and systematic samples: a maximum of 200 articles per UoA that were in most cases taken from the highest and lowest scoring departments.

Recall that the correlations are underestimates because of the use of a proxy research quality score. The results are also not directly comparable with the Cohen's kappa of the biomedical study because of the different metric, sampling method, and quality type (Wu et al., 2025). Nevertheless, the current results are more promising in the sense of covering more fields and giving stronger evidence of an ability to rank articles for research quality.

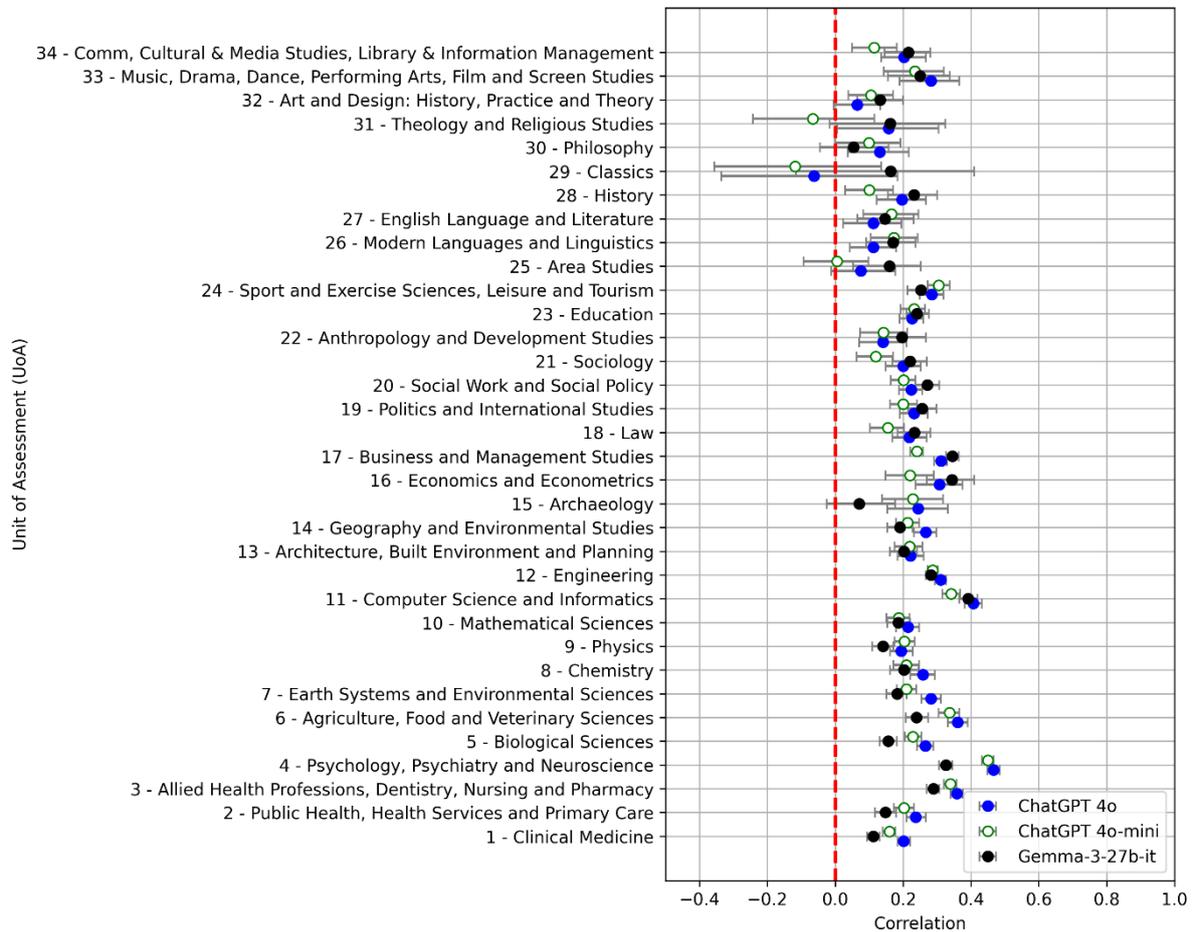

Figure 3. Spearman correlations for ChatGPT 4o, ChatGPT 4o-mini and Gemma-3-27b-it research quality scores against departmental average REF scores as a proxy for individual article scores, by UoA. The ChatGPT data is from (Thelwall, 2025b).

## Report structures

One obvious difference between ChatGPT 4o/4o-mini and Gemma 3 is that Gemma's reports tend to follow a standard structured format instead of the highly varied structures used by ChatGPT 4o and 4o-mini. For example, analysing the first set of reports for Main Panel D, the reports always started with a heading stating that the report was an assessment or evaluation of the article, giving its title. The second paragraph was then usually an overall score, and this is usually followed by a heading "Justification" followed by separate scores and justifications for Originality, Significance and Rigour (e.g., Table 2). The sample report in the appendix illustrates the standard structure of the reports.

Table 2. The contents of the first three paragraphs of the first set of Gemma-3-27b-it reports for Main Panel D (n=4966).

| Para. | Freq. | Content |
| --- | --- | --- |
| 1 | 4956 | "## Assessment of:", followed by title (with or without single/double quotes) |
| 1 | 10 | "## Evaluation of:", followed by title (with or without single/double quotes) |
| 2 | 3446 | Overall score, with or without a descriptor. |
| 2 | 772 | Explanation that an assessment of the article abstract follows |
| 2 | 748 | Explanation that an assessment of the article follows |
| 3 | 196 | The phrase "Detailed assessment" |
| 3 | 2840 | The phrase "Justification" |
| 3 | 331 | The phrase "Justification", followed by a justification |
| 3 | 273 | An originality score |
| 3 | 1248 | Overall score, with or without a descriptor. |
| 3 | 57 | Evaluative summary of the paper. |
| 3 | 16 | The phrase "Rationale", followed by a justification |
| 3 | 5 | Here's a detailed breakdown of the assessment based on the provided criteria: |

Analysing the common headings, including emphasised phrases at the start of paragraphs, virtually all reports contained separate sections for rigour, originality and significance and an overall score. The vast majority also contained a section justifying the score and a conclusion or summary (Table 3).

Table 3. The most frequent headings or starts of paragraphs in Gemma-3-27b-it reports for Main Panel D (n=24,830 reports).

| Heading/paragraph start | Frequency | Percentage |
|---|---|---|
| **3. Rigour… | 24827 | 100.0% |
| **1. Originality… | 24824 | 100.0% |
| **2. Significance… | 24822 | 100.0% |
| **Overall Score… | 24784 | 99.8% |
| **Justification… | 20848 | 84.0% |
| **In conclusion… | 19529 | 78.7% |
| ## Assessment of … | 12556 | 50.6% |
| **Detailed Breakdown… | 8327 | 33.5% |
| Here's an assessment of … | 7319 | 29.5% |

On the few occasions that an article was given different scores by Gemma, this was sometimes due to Gemma reporting an unrounded score in one report and only a rounded score in another (e.g., "**Overall Score: 2.5* (Rounded to 3*)**"). In very rare cases, different scores were given for the same criteria, as the two sections below about significance extracted from different reports for the same article illustrate. The reports are about a *New England Journal of Medicine* article with evidence from a clinical trial that a new treatment did not work. The second report includes additional negative points compared to the first one. Perhaps counterintuitively, however, Generative LLMs produce text with a probability model, token by token, so the justification is partly based on the score rather than the other way round. Thus, the additional criticisms are not the reason for the lower score, but the lower score could be at least partly the reason for the additional criticism. Nevertheless, both the lower score and the additional criticism could be at least partly due to the model attaching more importance to effect size when reading the abstract. If so, the explanation given in the paragraph would be helpful to understand the lower score. Unfortunately, it is not possible to test this.

> **2. Significance (3*):**
> [redacted summary of the study's significance] The study's results will likely influence clinical guidelines and future trial designs, potentially shifting focus towards alternative therapeutic strategies. The clear reporting of both primary and secondary outcomes, including adverse events, enhances its practical relevance. However, the lack of a positive finding limits its transformative impact on the field.

>  **Significance (2*)**
>
>  [redacted summary of the study's significance] This finding is important for guideline development and resource allocation. However, the effect size is small, and the confidence intervals are relatively wide, limiting the strength of the conclusions. The study's impact is more about *refining* understanding than dramatically altering it. While the findings will be noted and considered by clinicians, they are unlikely to fundamentally change the landscape of sepsis management.

The Gemma-3-27b-it are different from, and more standardised than, the reports from ChatGPT 4o for Main Panel D. For example, the common Gemma phrase, "Here's an assessment of" never occurs in ChatGPT 4o reports, and the closest, phrases starting, "Here is…" occur in less than 0.1% of ChatGPT 4o reports. At the simple level of characters/tokens, the common Gemma heading of "**1. Originality…" has many equivalents in ChatGPT 4o, including, "- **Originality:…", "### **Originality**:". "### **1. Originality**:…", and "- **Score for Originality:…". In terms of the overall structure, most ChatGPT 4o reports (4435/4966) start with phrases or headings meaning, "Evaluation of the article", but some reports start instead with a score, the article title, or a summary of the article.

## Conclusions

The results show, for the first time, that a downloadable open weights LLM can have a non-trivial ability to score academic journal articles for research quality, in the sense of producing scores that correlate positively with expert judgements. This capability exists for all or nearly all fields (not all correlations were statistically significantly different from zero). This means that downloadable LLMs can be used to support decision making in contexts where a quantitative indicator is needed, such as the situations were citation-based indicators are currently used. They can also be used for important research evaluations planned, such as the REF, where a guarantee of performance and system availability would be needed in advance. For a downloadable LLM there is no risk that the necessary capability disappears after a model is upgraded. Although the correlations are weaker than for ChatGPT 4o and ChatGPT 4o-mini, offline LLMs may be needed for high security contexts inside closed systems or for cost savings in some contexts. Switching from one of these models would give a small reduction in performance overall, with 83.8% of the correlation strength of ChatGPT 4o and 94.7% of the correlation strength of ChatGPT 4o-mini. Of course, since the inputs are titles and abstracts rather than full texts and LLMs are pattern matching and processing algorithms, there scores are not evaluations but are only guesses based on their world knowledge and the contents of each title and abstract.

At a more theoretical level, the results show that research quality scoring is not an emergent property of the largest current LLMs but also exists for the 27 billion parameter one used here. A 70 billion parameter version of the open weights LLM Llama 3 is sometimes used in research (e.g., Wu et al., 2025), so 27 billion is relatively modest in comparison. Nevertheless, there may still be a lower threshold, in terms of the number of parameters, below which research quality scoring does not give useful results. Of course, it may also not exist for other current LLMs of similar size or larger, but the main purpose of the current paper was to show that the capability can exist in open weights LLMs rather than to find the best one.

Finally, and at a more theoretical level, the scores and structures of the reports written by Gemma-3-27b-it are much less varied than those written by ChatGPT 4o-mini. Thus, report variety may be an emergent property of the largest current LLMs. Either their sizes or their architectures may allow them to exploit their knowledge in more different ways, perhaps releasing different research evaluation perspectives in their knowledge that allows their overall scores, after averaging multiple iterations, to be more powerful.

## Acknowledgements

This study is funded by the Economic and Social Research Council (ESRC), UK (APP43146).

## Appendix: Sample report (emphasis and redactions added)

## Assessment of: "[redacted]"
**Overall Score: 4***
**Justification:** This article represents world-leading research in its field, demonstrating exceptional quality across all three assessed criteria: originality, significance, and rigour. The discovery and comprehensive characterization of [redacted] is a landmark achievement with substantial implications for evolutionary biology and [redacted] research.

**1. Originality (4*):**
The discovery of [redacted] is a highly original contribution. While the search for such a species has been ongoing, its successful identification and detailed characterization represent a significant breakthrough. The research doesn't simply present new data; it *creates* a new system for investigation. The work goes beyond mere description, delving into genomic analysis, identifying unique features like the expansion of transposable elements and massive gene losses in [redacted]. This isn't just a descriptive genomic report; it's a comparative genomic analysis immediately framing evolutionary questions. The development of genetic and molecular techniques specifically for [redacted] further underscores the innovative nature of this work, providing the tools necessary for future research. The use of a novel ecological niche [redacted] as the source of the new species also demonstrates a creative approach to discovery.

**2. Significance (4*):**
The significance of this discovery is substantial. [redacted] is a cornerstone of biological research, and the availability of [redacted] opens up unprecedented opportunities for comparative evolutionary studies. The authors explicitly state this potential, and it is readily apparent. The ability to compare genomes and phenotypes will allow researchers to dissect the genetic basis of adaptation, identify conserved and divergent pathways, and gain deeper insights into the evolution of development, behaviour, and genome structure. The findings regarding transposable element expansion and chemoreceptor gene loss are particularly significant, offering potential explanations for species divergence and adaptation to specific environments. The work has the potential to influence a broad range of research areas, from molecular biology and genetics to ecology and evolutionary biology. The creation of a new model system will undoubtedly stimulate further research and generate new knowledge.

**3. Rigour (4*):**
The research demonstrates exceptional intellectual coherence and methodological rigour. The study employs a multi-faceted approach, integrating morphological, developmental, behavioural, and genomic analyses. The genome sequencing and assembly [redacted] are presented with sufficient detail to assess their quality. The identification of unique genomic features is supported by robust bioinformatics analyses. The authors clearly articulate the methods used and the rationale behind their experimental design. The comparison between [redacted] and [redacted] is systematic and well-documented. The development of genetic and molecular tools for the new species indicates a commitment to ensuring the reproducibility and accessibility of the research. The logical flow of the paper, from discovery to characterization to potential applications, is highly coherent and demonstrates a strong understanding of the field.

**Concluding Remarks:**
This article is a truly outstanding piece of research. It represents a significant advance in our understanding of nematode evolution and provides a valuable new resource for the scientific community. The combination of originality, significance, and rigour firmly places this work at the forefront of its field, deserving of the highest possible rating.